\documentclass[a4paper]{jpconf}
\usepackage{graphicx}
\usepackage{amssymb,amsfonts,amsmath,amsbsy,mathtext,cite,enumerate,float, bm}
\begin{document}
\title{Neutrino spin-flavor oscillations derived from the mass basis}

\author{R Fabbricatore$^1$, A Grigoriev$^2$, A Studenikin$^{1,3}$}

\address{${}^1$ Department of Theoretical Physics, Faculty of Physics, Lomonosov Moscow State University, Moscow, 119991, Russia}
\address{${}^2$ Skobeltsyn Institute of Nuclear Physics, Lomonosov Moscow State University, Moscow, 119991, Russia}
\address{${}^3$ Joint Institute for Nuclear Research, Dubna, Moscow Region, 141980, Russia}
\ead{studenik@srd.sinp.msu.ru}

\begin{abstract}

 We consider neutrino mixing and oscillations in presence of an arbitrary constant
magnetic field with nonzero transversal $B_{\perp}$ and longitudinal
$B_{\parallel}$ components with respect to the direction of neutrino propagation.
The electromagnetic interaction of neutrinos is determined by diagonal
and transition neutrino magnetic moments that are introduced for the neutrino mass states.
Explicit expressions for the effective neutrino diagonal and transition
magnetic moments for the flavor basis in terms of these values for the mass states
are obtained. The effective evolution Hamiltonian for the flavor neutrino and
the corresponding oscillation probability are derived. The role of the longitudinal magnetic field component  is examined. In particular, it is shown that:  1) $B_{\parallel}$ coupled to the corresponding magnetic
moments shifts the neutrino energy, and
2) in case of nonvanishing neutrino transition magnetic moments $B_{\parallel}$
produces an additional mixing between neutrino states, both in the mass and
flavor neutrino bases.

\end{abstract}

\section{Introduction}
The Nobel Prize in physics awarded in 2015 to
Takaaki Kajita and Arthur McDonald for studies of the atmospheric and solar neutrinos \cite{Fukuda:1998}
does not leave any doubt that neutrinos oscillate and have nonzero masses. The latter fact leads to the well-known possibility for neutrino to have non-trivial electromagnetic properties \cite{Fujikawa:1980yx}, which brought forth the research area that was investigated in details by numerous authors (see recent review  \cite{Giunti:2014ixa} and references therein). In the course of these studies many phenomena that may appear in electromagnetic fields have been recognized and described thoroughly. Among them the neutrino spin-flavor oscillations is the one, featuring both the above mentioned basic neutrino aspects -- nonzero mass and electromagnetic properties from the one side and mixing from another \cite{Cisneros:1971, Lihkachev:1995}. Owing to this, in spite of being a longstanding problem, the spin-flavor oscillations can reveal some new aspects of existence of neutrino mass and electromagnetic properties. These concern the problems of neutrino parameters relation in neutrino physical (mass) and flavor bases and, generally speaking, of an accurate derivation
of the formulas used to describe oscillations.

We below  give an advanced view on the standard scheme of neutrino spin-flavor oscillations description aiming at the solid determination of parameters involved in the formalism and its
rigorous derivations.  We consider neutrino mixing and oscillations in presence of an arbitrary constant
magnetic field with nonzero transversal $B_{\perp}$ and longitudinal
$B_{\parallel}$ components with respect to the direction of neutrino propagation.
The electromagnetic interaction of neutrinos is determined by neutrino diagonal
and transition magnetic moments that are introduced for the neutrino mass states.
Explicit expressions for the effective neutrino diagonal and transition
magnetic moments for the flavor bases in terms of these values for the mass states
are obtained. The effective evolution Hamiltonian for the flavor neutrino and
the corresponding oscillation probability are derived. The role the longitudinal magnetic field component  is examined. In particular, it is shown that:  1) $B_{\parallel}$ coupled to the corresponding magnetic
moments shifts the neutrino energy, and
2) in case of nonvanishing neutrino transition magnetic moments $B_{\parallel}$
produces an additional mixing between neutrino states, both in the mass and
flovour neutrino bases.

\section{Neutrino spin oscillations in mass basis}
Consider two Dirac neutrino physical states, $\nu_1$ and $\nu_2$,
with masses $m_1$ and $m_2$.
Also consider the neutrino electromagnetic interaction via magnetic moment matrix $\mu_{\alpha \beta}$ ($\alpha,  \beta=1,2$):
\begin{equation}\label{em_hamilt}
H_{EM} = \frac{1}{2}\mu_{\alpha \beta}\overline{\nu}_{\alpha}\sigma_{\mu \nu}\nu_{\beta}F^{\mu \nu} + h.c. \ ,
\end{equation}
where $F^{\mu \nu}$ is the electromagnetic field tensor, $\sigma_{\mu \nu}=i/2 (\gamma_{\mu}\gamma_{\nu}-\gamma_{\nu}\gamma_{\mu})$ and $\gamma_{\mu}$ being the Dirac matrices. In a uniform magnetic field the Hamiltonian (\ref{em_hamilt}) becomes
\begin{equation}\label{B_hamilt}
	H_{EM} = -\mu_{\alpha \alpha '}\overline{\nu}_{\alpha }\bm{\Sigma B}\nu_{\alpha '} + h.c. ,
\end{equation}
where
\begin{equation}
\label{Sigma}
\Sigma_i=\left(\begin{array}{cc}
                   \sigma_i & 0 \\
                   0 & \sigma_i
                 \end{array}\right),
\end{equation}
$\sigma_i$ are the Pauli matrices.

For considering the neutrino evolution in the ultrarelativistic limit we introduce the 4-component basis $(\nu_{1, s=1}, \nu_{1, s=-1}, \nu_{2, s=1}, \nu_{2, s=-1})$ of states with definite helicity $s=\pm 1$. With the standard column vector notation, $\nu_m \equiv (\nu_{1, s=1}, \nu_{1, s=-1}, \nu_{2, s=1}, \nu_{2, s=-1})^{T}$ the neutrino evolution equation relevant to electromagnetic interaction has the Schr$\ddot{o}$dinger-like form,
\begin{equation}\label{schred_eq}
	i\frac{d}{dt} \nu_m (t)=H_{eff}\nu_m (t).
\end{equation}
The effective Hamiltonian consists of the vacuum and interaction parts
\begin{equation}\label{H_eff}
  H_{eff}=H_{vac}+H_{B},
\end{equation}
where the interaction part $H_{B}$ is composed of matrix elements of the interaction Hamiltonian (\ref{B_hamilt}) taken over the helicity neutrino states: $H_B=\langle	 {\nu_{\alpha,s}}|H_{EM}|{\nu_{\alpha ', s'}}\rangle$.

Let us calculate the effective interaction Hamiltonian under the assumption that neutrino moves along the $z$-axis. From the magnetic field interaction Hamiltonian (\ref{B_hamilt}) we have:
\begin{equation}\label{HB}
H^B_{\alpha,s; \alpha ', s'}=\langle	{\nu_{\alpha,s}}|H_{EM}|{\nu_{\alpha ', s'}}\rangle=-\frac{\mu_{\alpha, \alpha '}}{2}\int{d^3x \nu_{\alpha}^{\dagger}\gamma_0\begin{pmatrix}\bm{\Sigma B} & 0 \\ 0 & \bm{\Sigma B}\end{pmatrix}\nu_{\alpha '}}.
\end{equation}
For the spinors representing the free neutrino states we take
\begin{equation}	 \label{wave func}
\nu_{\alpha,s}=C_{\alpha}\sqrt{\frac{E_{\alpha}+
m_{\alpha}}{2E_{\alpha}}}\begin{pmatrix}u_{s} \\ \frac{\bm{\Sigma p_{\alpha}}}{E_{\alpha}+m_{\alpha}}u_{s}\end{pmatrix}e^{i\bm{p_{\alpha} x}},
\end{equation}
where $\bm{p_{\alpha}}$ is the neutrino $\nu_{\alpha}$ momentum. The two-component spinors $u_s$ define neutrino helicity states, and are given by
\begin{equation}\label{u_s_plus}
	u_{s=1}=\begin{pmatrix}1 \\ 0\end{pmatrix}, \ \ \ \
	u_{s=-1}=\begin{pmatrix}0 \\ 1\end{pmatrix}.
\end{equation}
Recall that in the ultrarelativistic limit these two states correspond to the right-handed $\nu_{R}$ and left-handed $\nu_{L}$ chiral neutrinos, respectively.

Substituting (\ref{wave func}) into the effective Hamiltonian given by (\ref{HB}), we get
\begin{multline}\label{HB2} H^B_{\alpha,s; \alpha ', s'}=-\frac{1}{2}\mu_{{\alpha \alpha '}}
C_{\alpha}C_{\alpha '}\int{d^3x\bm{B}\begin{pmatrix}u_{s}^{\dagger}, & u_{s}^{\dagger}\frac{\bm{\Sigma p_{\alpha}}}{E_{\alpha}+m_{\alpha}}\end{pmatrix} \begin{pmatrix}\bm{\Sigma} & 0 \\ 0 & -\bm{\Sigma}\end{pmatrix}\begin{pmatrix}u_{s'} \\ \frac{\bm{\Sigma p_{\alpha '}}}{E_{\alpha '}+m_{\alpha '}}u_{s'}\end{pmatrix}}\\
\\ \times \frac{\sqrt{\left(E_{\alpha}+m_{\alpha}\right)\left(E_{\alpha '}+m_{\alpha '}\right)}}{2\sqrt{E_{\alpha}E_{\alpha '}}} \exp\left(i\Delta px\right) .\end{multline}
Decomposing the magnetic field vector into longitudinal and transversal with respect to neutrino motion components $\bm{B}=\bm{B_{||}}+\bm{B_{\perp}}$ it is possible to show that
\begin{multline}
\bm{B}\begin{pmatrix}u_{s}^{\dagger} & \frac{\bm{\Sigma p_{\alpha}}}{E_{\alpha}+m_{\alpha}}u_{s}^{\dagger}\end{pmatrix} \begin{pmatrix}\bm{\Sigma} & 0 \\ 0 & -\bm{\Sigma}\end{pmatrix}\begin{pmatrix}u_{s'} \\ \frac{\bm{\Sigma p_{\alpha '}}}{E_{\alpha '}+m_{\alpha '}}u_{s'}\end{pmatrix}=\\
\\u_{s}^{\dagger} \left( \bm{\Sigma B_{||}}\left(1-\frac{\bm{p_{\alpha} p_{\alpha '}}}{\left(E_{\alpha}+m_{\alpha}\right)\left(E_{\alpha '}+m_{\alpha '}\right)}\right)+\bm{\Sigma B_{\perp}}\left(1+\frac{\bm{p_{\alpha} p_{\alpha '}}}{\left(E_{\alpha}+m_{\alpha}\right)\left(E_{\alpha '}+m_{\alpha '}\right)} \right) \right)  u_{s'}.
\end{multline}
In the the ultrarelativistic limit $\frac{m_{\alpha}}{E_{\alpha}} \ll 1$ one gets:
\begin{equation}
\left(1-\frac{\bm{p_{\alpha} p_{\alpha '}}}{\left(E_{\alpha}+m_{\alpha}\right)\left(E_{\alpha '}+m_{\alpha '}\right)}\right)\frac{\sqrt{\left(E_{\alpha}+m_{\alpha}\right)\left(E_{\alpha '}+m_{\alpha '}\right)}}{2\sqrt{E_{\alpha}E_{\alpha '}}} \approx
\frac{1}{2}\left(\frac{m_{\alpha}}{E_{\alpha}}+\frac{m_{\alpha '}}{E_{\alpha '}}\right) \equiv \gamma^{-1}_{\alpha \alpha '},
\end{equation}
where the quantity $\gamma_{\alpha \alpha '}$, fetched out also in \cite{Akhmed:1988}, we call the transition gamma-factor. Similarly, it is also possible to show that
\begin{equation}\left(1+\frac{\bm{p_{\alpha} p_{\alpha '}}}{\left(E_{\alpha}+m_{\alpha}\right)\left(E_{\alpha '}+m_{\alpha '}\right)}\right)\frac{\sqrt{\left(E_{\alpha}+m_{\alpha}\right)\left(E_{\alpha '}+m_{\alpha '}\right)}}{2\sqrt{E_{\alpha}E_{\alpha '}}} \approx 1.\end{equation}

Introducing an angle $\beta$ between the $\bm{B}$ and $\bm{p_{\alpha}}$ vectors and assuming that $\bm{B_{\perp}}$ is aligned along the $x$-axis, we further obtain that
\begin{multline}
u^{\dagger}_{s=1}\bm{\Sigma B}u_{s=1}=\begin{pmatrix}1 & 0\end{pmatrix}\bm{\Sigma}\begin{pmatrix}1 \\ 0\end{pmatrix}\left(\bm{B_{||}}+\bm{B_{\perp}}\right)=\\\begin{pmatrix}1 & 0\end{pmatrix}\sigma_3\begin{pmatrix}1 \\ 0\end{pmatrix}B\cos \beta + \begin{pmatrix}1 & 0\end{pmatrix}\sigma_1\begin{pmatrix}1 \\ 0\end{pmatrix}B\sin \beta=B\cos \beta ,\end{multline}
and, similarly,
\begin{equation}u^{\dagger}_{s=1}\bm{\Sigma B}u_{s=-1}=B\sin \beta ,\end{equation}
\begin{equation}u^{\dagger}_{s=-1}\bm{\Sigma B}u_{s=1}=B\sin \beta ,\end{equation}
\begin{equation}u^{\dagger}_{s=-1}\bm{\Sigma B}u_{s=-1}=-B\cos \beta .\end{equation}
As it was expected, in neutrino transitions without change of helicity only the $B_{\parallel}=B\cos \beta$ component of the magnetic field contributes to the effective potential, whereas in transitions with change of the neutrino helicity the transversal component $B_{\perp}=B\sin \beta$ matters.

Performing the remaining simple algebra one can readily write out the $H_{B}$ matrix. Thus, we obtain the evolution equation (\ref{schred_eq}) in the following form
\begin{equation}\label{gen_evol_eq}
	i\frac{d}{dt} \begin{pmatrix}\nu_{1, s=1} \\ \nu_{1, s=-1} \\  \nu_{2, s=1 } \\ \nu_{2, s=-1 }\end{pmatrix}=
	\begin{pmatrix}
	E_{1}-\mu_{1 1}\frac{B_{||}}{\gamma_{1 1}} & \mu_{1 1}B_{\perp} & -\mu_{1 2 }
\frac{B_{||}}{\gamma_{1 2}} & \mu_{1 2 }B_{\perp} \\
	\mu_{1 1}B_{\perp} & E_{1}+\mu_{1 1}\frac{B_{||}}{\gamma_{1 1}} & \mu_{1 2}B_{\perp} &
\mu_{1 2}\frac{B_{||}}{\gamma_{1 2 }} \\
	-\mu_{1 2 }\frac{B_{||}}{\gamma_{1 2 }} & \mu_{1 2 }B_{\perp} & E_{2}-\mu_{2  2 }\frac{B_{||}}
{\gamma_{2  2 }} & \mu_{2  2 }B_{\perp} \\
	\mu_{1 2}B_{\perp} & \mu_{1 2}\frac{B_{||}}{\gamma_{1 2 }} & \mu_{2  2}B_{\perp} & E_{2}+
\mu_{2  2 }\frac{B_{||}}{\gamma_{2  2 }}\\
	\end{pmatrix}
	\begin{pmatrix}\nu_{1, s=1} \\ \nu_{1, s=-1} \\  \nu_{2, s=1 } \\ \nu_{2, s=-1}\end{pmatrix}.
\end{equation}
This equation governs all possible oscillations of the four neutrino states with defined masses ($m_1$ and $m_2$) and helicities
($s=1$ and $s=-1$) in the presence of a magnetic field.
The obtained expression (\ref{gen_evol_eq}) for the evolution Hamiltonian explicitly reveals a particular role
of the longitudinal magnetic field component  $B_{||}$:

1) the longitudinal magnetic field component  $B_{||}$  coupled to the corresponding magnetic moment shifts the neutrino energy,

2) in case of nonvanishing neutrino transition magnetic moment $\mu_{12}$, the presence of the longitudinal field $B_{||}$ produces mixing among neutrino species with different masses but with equal helicities.

At the same time, from (\ref{gen_evol_eq}) it is clearly seen that
mixings of different helicity neutrino states are
due to the corresponding magnetic moment (or transition magnetic moment)
interactions with the transversal magnetic field $B_{\perp}$.

\section{Effective neutrino magnetic moments in flavor basis}
Once having physics in the mass basis in hands, our next step is to bring it to observational terms. This means that we must elaborate a
generalization of the mixing matrix for transitions between neutrino state
vectors written in two four-component bases $\nu_m$ and $\nu_{f}\equiv (\nu_{e}^{R}, \nu_{e}^{L}, \nu_{\mu}^{R}, \nu_{\mu}^{L})^T$ so that
%\begin{equation}\label{U and neutrino vectros}
%\nu_{f}=
%\begin{pmatrix}
%\nu_{e}^{R}\\
%\nu_{e}^{L}\\
%\nu_{\mu}^{R}\\
%\nu_{\mu}^{L}\\
%\end{pmatrix},
%\
%\nu_m=
%\begin{pmatrix}
%\nu_{1,s=1}\\
%\nu_{1,s=-1}\\
%\nu_{2,s=1}\\
%\nu_{2,s=-1}\\
%\end{pmatrix}
%,
%\ \ \
%\nu_{f}=U\nu_m
%.
%\end{equation}
\begin{equation}\label{U_def}
  \nu_{f}=U\nu_m.
\end{equation}
This procedure appears to be not quite direct since we
should hold the condition that polarization of the fields
must be preserved under transformation of the bases elements. That is why, keeping in mind that chiral components are almost the helicity ones, we define
\begin{equation}\label{transformations}
    \begin{tabular}{ccc}
  $\nu_{e}^{R,L} =\nu_{1,s=\pm 1}\cos\theta+\nu_{2,s=\pm 1}\sin\theta$, \\
  $\nu_{\mu}^{R,L}=-\nu_{1,s=\pm 1}\sin\theta+\nu_{2,s=\pm 1}\cos\theta$.
  \end{tabular}
\end{equation}
Then, using Eqs. (\ref{U_def}) and (\ref{transformations}), it is easy to obtain that
\begin{equation}\label{U}
  U=
\begin{pmatrix}
    \cos\theta & 0 & \sin\theta & 0 \\
	0 & \cos\theta & 0 & \sin\theta \\
	-\sin\theta & 0 & \cos\theta & 0 \\
    0 & -\sin\theta & 0 & \cos\theta \\
\end{pmatrix}.
\end{equation}

Given the transition matrix (\ref{U}), derivation of the evolution equation in the flavor basis is straightforward:
%Leaving aside the oscillations due to kinetic part, which are are not the subject of our paper, we
\begin{equation}\label{schred_eq_fl}
  i\dfrac{d}{dt}\nu_{f}=UHU^{\dag}\nu_{f},
\end{equation}
so that the effective magnetic field interaction Hamiltonian
${\tilde H}_B^{f} \equiv U H_B U^{\dag}$ has the following structure,
%{\tiny
\begin{equation}\label{H_B_f}
{\tilde H}_B^{f}=
\begin{pmatrix}
{-\tilde{\mu}}_{ee}\frac{B_{||}}{\gamma_{ee}}  & \mu^{\prime}_{ee} B_{\perp} & -\tilde{\mu}_{e\mu} \frac{B_{||}}{\gamma_{e\mu}}  & \mu^{\prime}_{e\mu}B_{\perp}  \\
\mu^{\prime}_{ee} B_{\perp} & {\tilde{\mu}}_{ee} \frac{B_{||}}{\gamma_{ee}}  & \mu^{\prime}_{e\mu}B_{\perp} & \tilde{\mu}_{e\mu} \frac{B_{||}}{\gamma_{e\mu}}  \\
-\tilde{\mu}_{e\mu}\frac{B_{||}}{\gamma_{e\mu}}  & \mu^{\prime}_{e\mu}B_{\perp} & -{\tilde{\mu}}_{\mu\mu}\frac{B_{||}}{\gamma_{\mu\mu}}  & \mu^{\prime}_{\mu\mu} B_{\perp} \\
\mu^{\prime}_{e\mu}B_{\perp} & \tilde{\mu}_{e\mu}\frac{B_{||}}{\gamma_{e\mu}}  & \mu^{\prime}_{\mu\mu} B_{\perp} & {\tilde{\mu}}_{\mu\mu}\frac{B_{||}}{\gamma_{\mu\mu}}
\end{pmatrix}.
\end{equation}
%}
Here we have introduced the following formal notations intended to manifest an analogy with the standard spin-flavor oscillation formalism (see below):
\begin{eqnarray}\label{mu_fl}
  \mu^{\prime}_{ee} &=& \Big(\mu_{11}\cos^2\theta+\mu_{22}\sin^2\theta+\mu_{12}\sin2\theta\Big), \nonumber \\
  \mu^{\prime}_{e\mu} &=& \Big(\mu_{12}\cos2\theta+\frac{1}{2}(\mu_{22}-\mu_{11})\sin2\theta\Big), \\
  \mu^{\prime}_{\mu\mu} &=& \Big(\mu_{11}\cos^2\theta+\mu_{22}\sin^2\theta-\mu_{12}\sin2\theta\Big), \nonumber
\end{eqnarray}
\begin{eqnarray}\label{mu_Gammas_fl}
  \nonumber \frac{{\tilde{\mu}}_{ee}}{\gamma_{ee}} &=& \Big(\frac{\mu_{11}}{\gamma_{11}}\cos^2\theta+\frac{\mu_{22}}{\gamma_{22}}\sin^2\theta+\frac{\mu_{12}}{\gamma_{12}}\sin2\theta \Big), \\
  \frac{\tilde{\mu}_{e\mu}}{\gamma_{e\mu}} &=& \Big(\frac{\mu_{12}}{\gamma_{12}}\cos2\theta+\frac{1}{2}\Big(\frac{\mu_{22}}{\gamma_{22}}-\frac{\mu_{11}}{\gamma_{11}}\Big)
  \sin2\theta\Big), \\
  \nonumber \frac{{\tilde{\mu}}_{\mu\mu}}{\gamma_{\mu\mu}} &=& \Big(\frac{\mu_{11}}{\gamma_{11}}\cos^2\theta+\frac{\mu_{22}}{\gamma_{22}}\sin^2\theta-\frac{\mu_{12}}
  {\gamma_{12}}\sin2\theta\Big).
\end{eqnarray}

The Hamiltonian  (\ref{H_B_f}) has been derived within much consistent procedure
than is usually given in literature. It should be noted that  Eqs.~(\ref{mu_fl}) could be also obtained
as results of general consideration \cite{Giunti:2014ixa} of neutrino mixing and oscillations
in an arbitrary constant magnetic field
(the final form like that of Eqs.~(\ref{mu_fl})
was not established in \cite{Giunti:2014ixa}). Also, an effective neutrino oscillation Hamiltonian with terms containing quantities (\ref{mu_Gammas_fl}), however without account for possibility of spin transitions, was obtained in \cite{Akhmed:1988}.

\section{Discussion}
Let us now confront the obtained effective magnetic field interaction Hamiltonian (\ref{H_B_f}) with the one typically written straight in the neutrino flavor basis (see, for instance, \cite{Giunti:2014ixa} and \cite{Lihkachev:1995}):
\begin{equation}\label{H_B_f_old}
H_B^{f}=
\begin{pmatrix}
-{{\mu}}_{ee}\frac{B_{||}}{\gamma}  & \mu_{ee} B_{\perp} & 0  & \mu_{e\mu}B_{\perp} \\
\mu_{ee} B_{\perp} & {\mu}_{ee} \frac{B_{||}}{\gamma}  & \mu_{e\mu}B_{\perp} & 0 \\
0 & \mu_{e\mu}B_{\perp} & -{\mu}_{\mu\mu}\frac{B_{||}}{\gamma}  & \mu_{\mu\mu} B_{\perp} \\
\mu_{e\mu}B_{\perp} & 0 & \mu_{\mu\mu} B_{\perp} & {\mu}_{\mu\mu}\frac{B_{||}}{\gamma}
\end{pmatrix},
\end{equation}
where $\gamma$ is the common neutrino gamma-factor.

First of all, we can ascertain that the structure of the obtained expression (\ref{H_B_f}) is consistent with the ``standard" Hamiltonian (\ref{H_B_f_old}). At that, the magnitudes (\ref{mu_fl}) account for the neutrino mixings and represent the straightforward expressions for the effective magnetic moments in the flavor basis in terms of magnetic moments introduced in the mass basis. The neutrino effective ``magnetic moments" (\ref{mu_Gammas_fl}) determine neutrino interactions and mixings due to the longitudinal magnetic field $B_{||}$.

As in the mass basis, the magnetic field component  $B_{||}$  shifts the neutrino energy. Also, it contributes to transitions with change of flavor. It is interesting to observe that when the transversal field component is set to zero the Hamiltonian structure becomes
\begin{equation}\label{H_flavor_decoupled}
\begin{pmatrix}
-a  & 0 & -b  & 0 \\
0 & a & 0 & b \\
-b & 0 & -c  & 0 \\
0 & b & 0 & c
\end{pmatrix},
\end{equation}
so that neutrino states with different flavors and the same chirality decouple and form subsystems independently mixed by the field.
For example, one would have two neutrino species $\nu^L_{e}$ and $\nu^L_{\mu}$ mixed in accordance with the equation (here the standard terms for vacuum oscillations are added)
\begin{equation}\label{3_evol_eq}
	i\frac{d}{dt} \begin{pmatrix}\nu^L_{e} \\ \nu^L_{\mu} \\  \end{pmatrix}=
	\begin{pmatrix}
	{-\frac{\Delta m^2}{4 E}\cos 2\theta+\tilde{\mu}}_{ee}\frac{B_{||}}{\gamma_{ee}} &  \frac{\Delta m^2}{4 E}\sin 2\theta+\tilde{\mu}_{e\mu} \frac{B_{||}}{\gamma_{e\mu}}  \\
	 \frac{\Delta m^2}{4 E}\sin 2\theta+\tilde{\mu}_{e\mu} \frac{B_{||}}{\gamma_{e\mu}} & \frac{\Delta m^2}{4 E}\cos 2\theta + {\tilde{\mu}}_{\mu\mu}\frac{B_{||}}{\gamma_{\mu\mu}}  \\
		\end{pmatrix}
	\begin{pmatrix}\nu^L_{e} \\ \nu^L_{\mu} \\ \end{pmatrix}.
\end{equation}
In this way, neutrino oscillations would be influenced by the longitudinal field $B_{||}$ (see also \cite{Akhmed:1988}) and   $B_{||}$ would produce an additional mixing with respect to the usual one described by the vacuum mixing angle $\theta$.

The oscillation probability is just strightforward,
\begin{equation}\label{P_oscillation}
  P_{\nu^L_{e}\rightarrow \nu^L_{\mu}}
     = \frac{
             \left(
                  \frac{\Delta m^2}{2 E}\sin 2\theta+2\tilde{\mu}_{e\mu} \frac{B_{||}}{\gamma_{e\mu}}
             \right)^2
             }
            {
             \left(
                  \frac{\Delta m^2}{2 E}\sin 2\theta+2\tilde{\mu}_{e\mu} \frac{B_{||}}{\gamma_{e\mu}}
             \right)^2
                   +
             \left(
                  \frac{\Delta m^2}{2 E}\cos 2\theta - 2 {\mu}_{12}\frac{B_{||}}{\gamma_{12}} \sin 2\theta
             \right)^2
            }
            \sin^2 \left( \frac{1}{2}\sqrt{D}x \right),
\end{equation}
where $D$ is the denominator of the pre-sine factor and the other quantities have usual for the theory of neutrino oscillations meaning. It follows that in case the second
term in the denominator is much smaller than the first one then the
amplitude of the flavor oscillations
$\nu^L_{e} \Leftrightarrow\nu^L_{\mu}$ gets its maximal value. This can
be considered as an effect of neutrinos interaction with
the longitudinal magnetic field due to neutrino magnetic moments.

\ack
One of the authors (A.S.) is thankful to Nicolao Fornengo and Carlo Giunti for the kind invitation to
participate in the 14th International Conference on Topics in Astroparticle and Underground
Physics.   The  work  on  this  paper  was  partially  supported  by  the  Russian  Basic  Research
Foundation grants no. 14-22-03043-ofi-m, 15-52-53112-gfen and 16-02-01023-a.


\section*{References}
\begin{thebibliography}{4}
\bibitem{Fukuda:1998}
  Fukuda Y {\it et al} (Super-Kamiokande Collaboration)
  1998
  %``Neutrino electromagnetic interactions: a window to new physics,''
  {\it Phys. Rev. Lett.} {\bf 81} 1562
\nonum
  Ahmad Q R {\it et al} (SNO collaboration)
  2001
  %``Neutrino electromagnetic interactions: a window to new physics,''
  {\it Phys. Rev. Lett.} {\bf 87} 071301
\nonum
  Ahmad Q R {\it et al} (SNO collaboration)
  2002
  %``Neutrino electromagnetic interactions: a window to new physics,''
  {\it Phys. Rev. Lett.} {\bf 89} 011301
\bibitem{Fujikawa:1980yx}
Fujikawa K and Shrock R
1980
%"The Magnetic Moment of a Massive Neutrino and Neutrino Spin Rotation",
{\it Phys. Rev. Lett.} {\bf 45} 963
\bibitem{Giunti:2014ixa}
  Giunti C and Studenikin A
  2015
  %``Neutrino electromagnetic interactions: a window to new physics,''
  {\it Rev. Mod. Phys.}  {\bf 87} 531
  %doi:10.1103/RevModPhys.87.531
  ({\it Preprint} arXiv:1403.6344 [hep-ph])
\bibitem{Cisneros:1971}
  Cisneros A
  1971
  {\it Astrophys. Space Sci.} {\bf 10} 87
\nonum
  Schechter J and  Valle J
  1981
  {\it  Phys.  Rev.}  {\bf D} 24  1883
\nonum
  Voloshin M B, Vysotsky M I and  Okun L B
  1986
  {\it Sov. Phvs.  JETP}  {\bf 64} 446
\nonum
  Akhmedov E K
  1988
  {\it Sov. J. Nucl. Phys.} {\bf 48} 382
\nonum
  Lim  C-S and Marciano W J
  1988
  {\it Phys. Rev.} D {\bf 37} 1368
\bibitem{Lihkachev:1995}
  Likhachev G  and Studenikin A
  1995
  {\it Sov. Phys.  JETP} {\bf 81} 419
\bibitem{Akhmed:1988}
  Akhmedov E K and Khlopov M Yu
  1988
  {\it Sov. J. Nucl. Phys.} {\bf 47} 689-691
\nonum
  Akhmedov E K and Khlopov M Yu
  1988
  {\it Mod. Phys. Lett.} {\bf A3} 451-457
\end{thebibliography}
\end{document}